\documentclass[preprint,aps]{revtex4}
\usepackage{graphicx}
\usepackage{bm}

\begin{document}
\title{Collision of spinning black holes in the close limit: The parallel spin case}
\author{Gaurav Khanna}
\affiliation{Theoretical and Computational Studies Group, \\
Southampton College of Long Island University,\\
Southampton NY 11968.}

\date{\today}

\begin{abstract}
In this paper we consider the collision of black holes with parallel spins using 
first order perturbation theory of rotating black holes (Teukolsky formalism).  
The black holes are assumed to be close to each other, initially non boosted and 
spinning slowly. We estimate the properties of the gravitational radiation released 
from such an collision. The same problem was studied recently by Gleiser {\em et al.} 
in the context of the Zerilli perturbation formalism and our results for waveforms, 
energy and angular momentum radiated agree very well with the results presented in 
that work.    
\end{abstract}

\maketitle

\section{Introduction}

There is considerable current interest in studying the collision of
two black holes, since these events could be primary sources of
gravitational waves for interferometric gravitational wave detectors
currently under construction. The close limit approximation applies to the 
late stage of such an event, when the system can be considered as a 
single distorted black hole. One can then use linear perturbation theory  to
estimate the energy and angular momentum lost by the system due to gravity 
wave emission, and even obtain ``waveforms'' which could be useful for 
experimental detection of these waves. 

In the past, this method has been applied to the case of a head-on collision 
of two boosted holes \cite{boost} and the case of slow inspiral \cite{njp} with 
considerable success. Recently, this method was extended to cover the
collision of spinning holes with anti-parallel spin \cite{isbh} and case of equal 
and parallel spin \cite{zerilli}. However in the latter work, the close limit of 
the two merging holes was considered as a distorted Schwartzchild hole. Therefore, 
the perturbation formalism used in that work, was the Zerilli formalism. In this 
paper, we shall treat the merger of two equal mass holes with equal and parallel 
spin, as a distorted Kerr black hole. Thus, we shall use the Teukolsky 
formalism for our evolutions, and compare our results with those from the Zerilli 
formalism.  

We also hope that such a comparative study, will shed light on the apparent 
discrepancy that was noted for the amount of the angular momentum lost for the 
case of  slow inspiral of two equal mass holes, as obtained by these two perturbation 
formalisms  \cite{njp}. Comments relating to that shall be published elsewhere. 

It should be noted that, results presented here can be easily combined with results 
obtained in the past, (say)  for the slow inspiral case \cite{njp} via simple superposition, to obtain waveforms, etc. for an event in which two equal mass and equal and parallel spin holes merge.  This is because all these results are based on linear black hole perturbation theory .

\section{Initial data}

To evolve a spacetime in general relativity, one needs as 
initial data, a 3-geometry $g_{ab}$ and an extrinsic curvature
$K_{ab}$, that solve Einstein's equations on some starting
hypersurface. These initial value equations have the form,
\begin{eqnarray}
\nabla^a (K_{ab} - g_{ab} K) &=& 0\\
{}^3R-K_{ab} K^{ab} + K^2 &=&0
\end{eqnarray}
where ${}^3R$ is the scalar curvature of the three metric. If
we propose a 3-metric that is conformally flat $g_{ab} = \phi^4
\widehat{g}_{ab}$, with $\widehat{g}_{ab}$ a flat metric, and $\phi^4$
the conformal factor, and we use a decomposition of the extrinsic
curvature $K_{ab} = \phi^{-2} \widehat{K}_{ab}$, and assume maximal
slicing $K_a^a=0$, the constraints become \cite{BoYo},
\begin{eqnarray}
\widehat{\nabla}^a \widehat{K}_{ab} &=& 0\label{momentum}\\
\widehat{\nabla}^2 \phi &=& -\frac{1}{8}
\phi^{-7} \widehat{K}_{ab} \widehat{K}^{ab}\ ,\label{hami}
\end{eqnarray}
where $\widehat{\nabla}$ is a flat-space covariant derivative.

To solve the momentum constraint, we start with the well known solution 
that represents a single hole with spin $S$,
\begin{equation}
\hat{K}^{\rm one}_{ab} = {3 \over R^3} \left[{\epsilon_{cad}} {S^d}{n^c}{n_b} 
+{\epsilon_{cbd}}{S^d}{n^c}{n_a}\right]\ .
\end{equation}
In this expression for the conformally related extrinsic curvature at
some point $x^a$, the quantity $n_b$ is a unit vector, in the 
flat space with metric $\hat{g}_{ab}$, directed from a point
representing the location of the hole to the point $x^a$. The symbol
$R$ represents the distance, in the flat base space, from the point
of the hole to $x^a$. It is straightforward to show that 
the solution of the Hamiltonian constraint corresponding to equation (5)
corresponds to a spacetime with  ADM angular momentum $S^{a}$.

The next step is to modify this to represent holes centered at 
$x=\pm L/2$ in the  flat metric. Since the 
momentum constraint is linear, we can simply add two expressions
of the above form,
\begin{equation}
\label{two}
\widehat{K}^{\rm two}_{ab} =
\widehat{K}^{\rm one}_{ab}\left(x \rightarrow x-L/2,S_{z} = S \right)+
\widehat{K}^{\rm one}_{ab}\left(x \rightarrow x+
L/2,{S_{z}} =S \right)\ .
\end{equation}
We will now use a polar coordinate system in the flat
space determined by $\hat{g}_{ab}$ centered in the mid-point separating the 
two holes and label the polar coordinates as $(R,\theta,\varphi)$. Thus, $R$ will 
be the distance in the flat space from the midpoint between the holes.

To solve the Hamiltonian constraint \ref{hami}, we use the ``punctures'' anzatz, 
i.e. we assume that $\phi$ is of the form,
\begin{equation}
\label{punkt1}
\phi = \phi_{BL} +\phi_{Reg}
\end{equation}
where,
\begin{equation}
\label{punkt2}
\phi_{BL}= 1 + {M \over 2 R_1} + {M \over 2 R_2}
\end{equation}
 that is, $\phi_{BL}$ is taken as the Brill-Lindquist conformal factor
\cite{brilin}, and we demand that $\phi_{Reg}$ must be regular in the whole
conformal plane, and vanish for large $R_i$. Here, $R_1$ and $R_2$ indicate 
the distance measured in conformal space, from the each of the two holes. If $\phi$ 
is to be a solution of equation \ref{hami},  $\phi_{Reg}$ must satisfy the equation,
\begin{equation}
\label{punkt3}
\nabla^2 \phi_{Reg} = - {1 \over 8} {\widehat{K}^{ab}\widehat{K}_{ab} \over 
( \phi_{BL} +\phi_{Reg})^7}
\end{equation}
where $ \widehat{K}^{ab}$ is equal to $\widehat{K}^{\rm two}_{ab}$, as defined in 
equation \ref{two}.

It is straightforward to see that the total initial $\widehat{K}^{ab}$ scales linearly with $S$. 
 If we assume  that $S$ and $L$ are small quantites (the holes are close to each other 
initially, and spinning slowly) and are of the same order,  we need to keep 
only the quadratic terms in $S$ in the the source term of the Hamiltonian constraint,
and we can neglect the rest. Then the Hamiltonian constraint would become, 
\begin{equation}
\label{punkt5}
\nabla^2 \phi_{Reg} = - 288 J^2 {\sin^2(\theta) R\over (2 R+M)^7}
\end{equation}
where, $J=2S$ the total angular momentum of the system. We solve this equation,
following Gleiser {\em et al.} \cite{zerilli} and obtain the final result for $\phi$ accurate to
order $J^2$,
\begin{equation}
\label{punkt6}
 \phi_{Reg} =  \phi_{(0,0)}(R) Y_0^0(\theta,\varphi)+\phi_{(2,0)}(R) 
Y_2^0(\theta,\varphi)
\end{equation}
where
\begin{eqnarray}
\label{punkt8}
\Phi_{(0,0)}(R)  & = &  { 4 \sqrt{\pi} J^2 (M^4+10 R M^3
+40 R^2 M^2+40R^3 M+ 16R^4) \over 5 M^3(2 R+M)^5}
  \nonumber \\
 \Phi_{(2,0)}(R)  & = & - {32 \sqrt{ 5 \pi} J^2 R^2 
\over 25 M (2 R+M)^5}      
\end{eqnarray}
For the sake of completion we also list here explicitly, the components of extrinsic 
curvature, keeping only the lowest two orders,
\begin{eqnarray}
\label{Khat1}
\widehat{K}_{RR} & = &  -3 {J L^2 \sin(2 \varphi) \sin^2(\theta) \over 
R^5} \nonumber \\
\widehat{K}_{R \theta} & = & -{9 \over 8} { J L^2 \sin(2 \varphi) \sin
(\theta)  \cos(\theta) \over R^4} \nonumber \\
\widehat{K}_{R \varphi} & = & 3 {J \sin^2(\theta) \over R^2} 
- {3 \over 8}{J L^2 (7 + \cos^2(\varphi)(25 \cos^2(\theta)-19)) \sin^2(\theta)
\over R^4} \nonumber \\ 
\widehat{K}_{\theta \theta} & = & {3 \over 4} {J L^2 \sin(2 \varphi)
\cos^2(\theta)/R^3} \nonumber \\ 
\widehat{K}_{\theta \varphi} & = & - {3 \over 4} {J L^2
\sin(\theta) \cos(\theta) (1 +3 \cos^2(\varphi)-5 \cos^2(\varphi) \cos^2(\theta))
\over R^3} \nonumber \\ 
\widehat{K}_{\varphi \varphi} & = & {3 \over 4} {J L^2
\sin(2\varphi)(4-9 \cos^2(\theta)+5 \cos^4(\theta) )\over R^3}    
\end{eqnarray}

We must now map the coordinates of the initial value solution to the
coordinates for a Kerr black hole with $a=J/M$ background. To do this, 
we interpret the $R$  as the isotropic radial coordinate  and we relate 
it to the usual Boyer-Linquist radial coordinate $r$ by
$R=(\sqrt{r}+\sqrt{r-2M})^2/4$.  From this we arrive at the final perturbative 
expressions for the components of the metric and extrinsic curvature. 
We shall interpret these quantities as perturbations over a Kerr solution 
background with perturbation parameter $L^2$. It is important to note that
we computed the initial data only to order $J^2$ mainly for ease 
in analytic computation, and we do not interpret $J$ as a formal perturbation 
parameter. Since we are using the Teukolsky formalism for our evolutions,
$J$ is simply a background quantity and not a perturbation parameter. This 
is an important difference from the corresponding Zerilli computation, where
$J$ is a perturbation parameter on equal footing with $L$. 

Using those expressions for the metric and extrinsic curvature, we calculate 
initial data for the Teukolsky function $\Psi$ following the prescription provided 
in reference \cite{teukid}.  The expressions we arrived at, are far too long and 
complicated to include in this paper, and to be of any direct use. Explicit algebraic 
expressions for this initial data shall be provided (in the computer algebra system, 
MAPLE format or as FORTRAN code), on request,  to anyone interested.

\section{Evolution of the Data using the Teukolsky equation}

Given the Cauchy data from the last section, the time evolution is
obtained from the Teukolsky equation \cite{Te},
\begin{eqnarray}
&&
\Biggr\{\left[a^2\sin^2\theta-\frac{(r^2 + a^2)^2}{\Delta}\right]
\partial_{tt}-
\frac{4 M a r}{\Delta}\partial_{t\varphi}
+ 4\left[r+ia\cos\theta-\frac{M(r^2-a^2)}{\Delta}\right]\partial_t
\nonumber\\
&&+\,\Delta^{2}\partial_r\left(\Delta^{-1}\partial_r\right)
+\frac{1}{\sin\theta}\partial_\theta\left(\sin\theta\partial_\theta\right)
+\left[\frac{1}{\sin^2\theta}-\frac{a^2}{\Delta}\right]
\partial_{\varphi\varphi}\\
&&-\, 4 \left[\frac{a (r-M)}{\Delta} + \frac{i \cos\theta}{\sin^2\theta}
\right] \partial_\varphi
-\left(4 \cot^2\theta +2 \right)\Biggr\}\Psi=0,
\end{eqnarray}
where $M$ is the mass of the black hole, $a$ its angular momentum per
unit mass, $\Sigma\equiv 
r^2+a^2\cos^2\theta$, and $\Delta\equiv
r^2-2Mr+a^2$. Evolving the initial data we just calculated, with this equation
will enable us to extract gravity wave waveforms that correspond to the late 
stage merger of spinning holes. We use the $2+1$ dimensional Teukolsky evolution 
code written by Krivan {\em et al.} \cite{ntc} using a radial (tortoise coordinate 
grid) resolution of $M/20$ and an angular resolution of $\pi/40$. We can also 
estimate the energy carried away by these gravitational waves using  
\cite{CaLo},
\begin{equation}
\frac{dE}{dt}=\lim_{r\to\infty}\left\{ \frac{1}{4\pi r^{6}}
\int_{\Omega}d\Omega\left| \int_{-\infty}^{t}d\tilde{t}\ \Psi(\tilde{t
},r,\theta,\varphi) \right|^2\right\}, \quad
d\Omega=\sin\theta\ d\vartheta\ d\varphi.
\end{equation}
The angular momentum radiated  can similarly be calculated using  \cite{CaLo},
\begin{equation}
\frac{dJ_z}{dt}=-\lim_{r\to\infty}\left\{ \frac{1}{4\pi r^{6}}{\rm Re}\left[
\int_\Omega d\Omega
\left(\partial_\varphi\int_{-\infty}^{t}d\tilde{t}\
\Psi(\tilde{t},r,\theta,\varphi) \right)
\left(\int_{-\infty}^{t}dt^\prime\int_{-\infty}^{t^\prime}d\tilde{t}\ 
\overline{\Psi}(\tilde{t},r,\theta,\varphi)\right)\right]
\right\}.  
\end{equation}

\section{Results of the evolutions}

In this section we show waveforms and plots for energy and angular momentum 
radiated from the collision of two spinning holes (with parallel spin) and compare
the results to those obtained from the Zerilli formalism. Recall that the two
holes have equal mass and equal and parallel spin. 

The waveforms that follow, are for a collision of two
black holes that were initially separated by a conformal distance of
$1.0$ and have an individual spin of $0.05$ in units of ADM
mass. The waves were extracted at radial location with $r^{*}=25M$ and 
at a polar angle $\theta=\pi/2$.

In figure 1 we show the $m=2$ mode of the Teukolsky function as a 
function of time. We see the typical quasi-normal ringing, in both the real and imaginary parts 
of the function. Note that the imaginary part the of waveform, there appears to be
a mixing of frequencies. This is exactly what was observed by Gleiser {\em et al.} 
\cite{zerilli}. They noted a mixing of $\ell=2$ and $\ell=3$ spherical harmonic modes in 
their evolutions. This is the type of signal that gravity wave observatories like 
LIGO \cite{ligo}, will detect if they happen to witness a collision of the kind we are 
considering. 

Let us now turn to the results for the radiated energies. Figure 2 shows 
the radiated energy as a function of the initial total spin, for a fixed separation of the 
holes. Note that the Zerilli results agree very well upto about $J/M^2=0.2$. Beyond that,
both these calculations cannot expect to yield accurate results. This is because both the 
methods are based on the close, slow approximation. Similarly, our results for radiated 
angular momentum, figure 3, agree very well to about $J/M^2=0.2$ but then diverge 
beyond that. 

One may make the observation that in both these cases (for large $a$) the Teukolsky formalism results appear to have more radiation than the corresponding Zerilli based results. This can be explained, in part, by the fact that there is less damping in the QN modes of a Kerr hole. However, it is  not clear that this physical reasoning accounts for the larger values completely. One must keep in mind, that for larger values of $a$, both these perturbative calculations break down, and therefore it is difficult to make any statements apart from observing some rather general trends. 

Also worth noting is that our results suggest that such a collision is unlikely to change 
the inspiral based estimate \cite{njp} of $1\%$ of the total system mass being radiated away by 
gravitational radiation. This is because, for every value of J (especially the larger values), 
this collision radiates much less than the inspiral case. One would therefore conclude that 
the radiation from the collision of two black holes is dominated by radiation coming from the
``inspiral'' part \cite{njp}. This fact was also observed in the context of holes with anti-parallel spins \cite{isbh}.   There also appears to be no appreciable change in the estimate for the radiated angular momentum. 

\section{A  ``realistic'' model}

In this work we have treated the collision of two black holes, with like masses and spins. A  more astrophysically likely event would be one that also has some orbital angular momentum, i.e. in addition to the two holes spinning, they are also inspiralling into each other.  In this section we shall attempt such an evolution. 

As mentioned before in this work, our approach here is going to be to use simple superposition of waveforms from our earlier work  \cite{njp} with the ones presented here. We can do this, since all these results are based on first-order perturbation theory of black holes. In addition to obtaining waveforms, we shall also obtain amounts for energy and angular momentum radiated. We choose two equal mass black holes located at $x =\pm 0.4$, with equal and parallel spins of $0.1$ each. The spins are aligned along the $z$-axis. Next we boost the black holes along the positive and negative $y$-axis respectively, so as to obtain a total orbital angular momentum of  $0.2$. Thus, the total angular momentum amounts to $0.4$. Please note that all the above mentioned quantities are in units of ADM mass. 

In Figure 4 we include waveforms from such an evolution. The figure depicts the real and imaginary parts of the $m=2$ mode of the Teukolsky function. It should be noted that this waveform is very close to the waveform produced by the ``inspiral'' part of this collision, indicating that, that is dominant  part of such an event. We also obtain estimates for energy and angular momentum carried away by these waves,
$$E_{RAD}/{M_{ADM}} = 1.0{\times}10^{-3} $$
$$J_{RAD}/{M_{ADM}}^{2} = 2.2 {\times}10^{-3}.$$
Moreover, by taking a difference between the energies obtained for a case in which the spins are aligned with the orbital angular momentum and the case in which they are anti-aligned, we can even estimate the lowest order coupling term between the orbital part and spin part. Also, by examining the analytic expressions for initial data, one can easily see that the form of this term has to be of the kind, $JSL^{2}$. Here $J$ is the orbital angular momentum, $S$ is the total spin and $L$ is the separation between the two holes. Numerical evolutions confirm the above and yield an empirical estimate for this term,
$$E_{CPLG}/{M_{ADM}} = (1.6{\times}10^{-3}) JSL^{2}/{M_{ADM}}^{6}.$$

\section{Conclusions}

We performed a first order, perturbative calculation to study the merger of
two spinning holes with parallel spin based on the Teukolsky formalism. Our results agree
very well with those obtained in the same context using the Zerilli formalism. 

We also noted that our results here do not significantly affect the original inspiral based 
estimates \cite{njp} for energy and angular momentum radiated.

\begin{figure*}
\includegraphics[height=10cm,width=16cm]{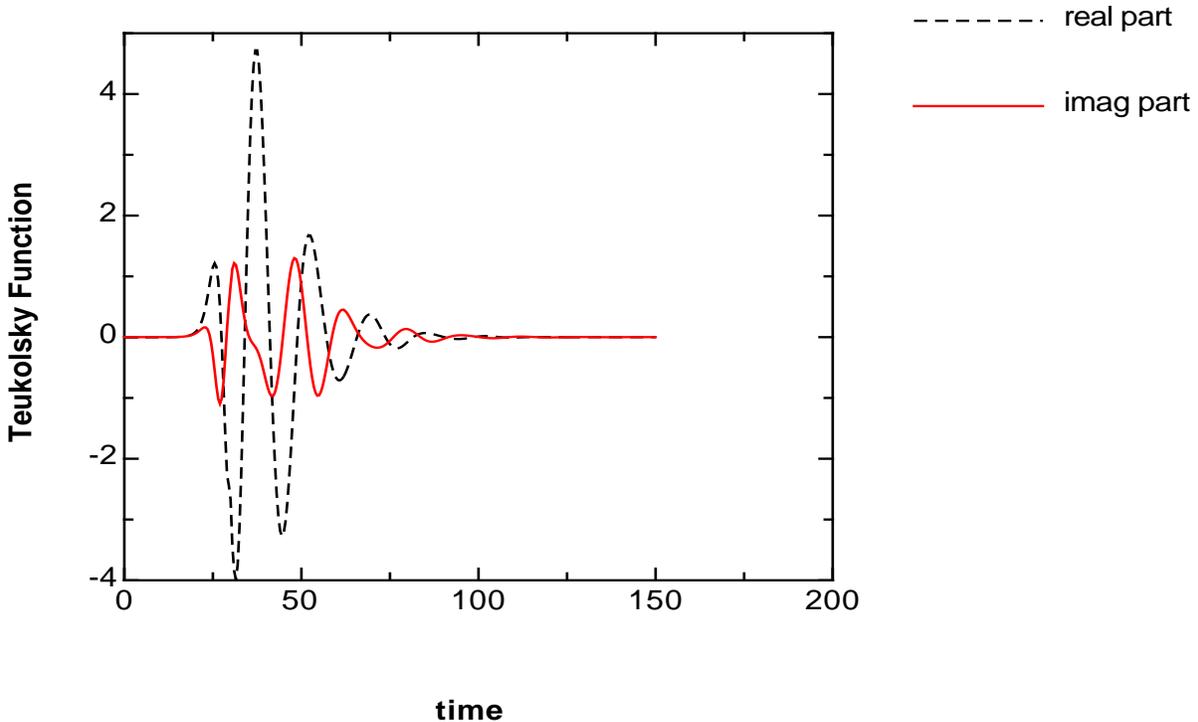}% Here is how to import EPS art
\label{waveform}
\caption{  Real and imaginary parts of the Teukolsky waveform extracted at $r^{*}=25M$. All 
quantities are in units of ADM mass.}
\end{figure*}

\begin{figure*}
\includegraphics[height=10cm,width=16cm]{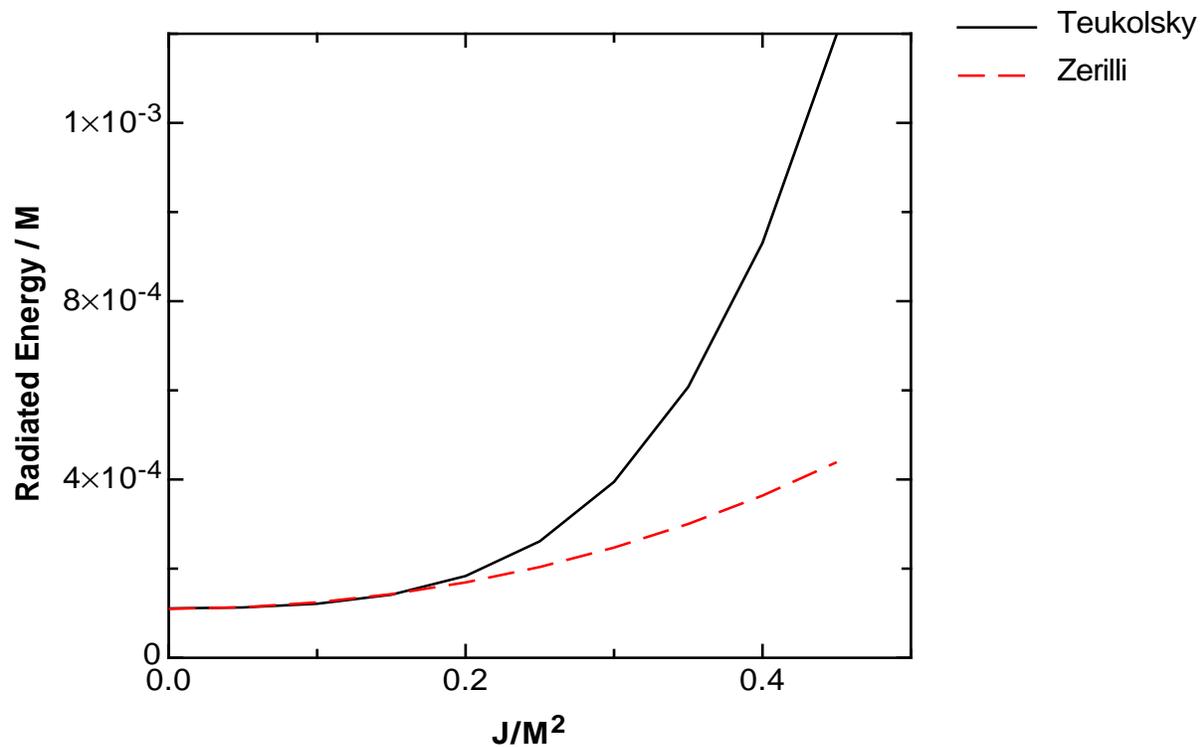}% Here is how to import EPS art
\label{energy}
\caption{ Radiated energies as a function of total initial angular momentum extracted at $r^{*}=25M$. Note the good agreement between the two formalisms upto $J/M^{2}=0.2$.
All quantities are in units of ADM mass.}
\end{figure*}

\begin{figure*}
\includegraphics[height=10cm,width=16cm]{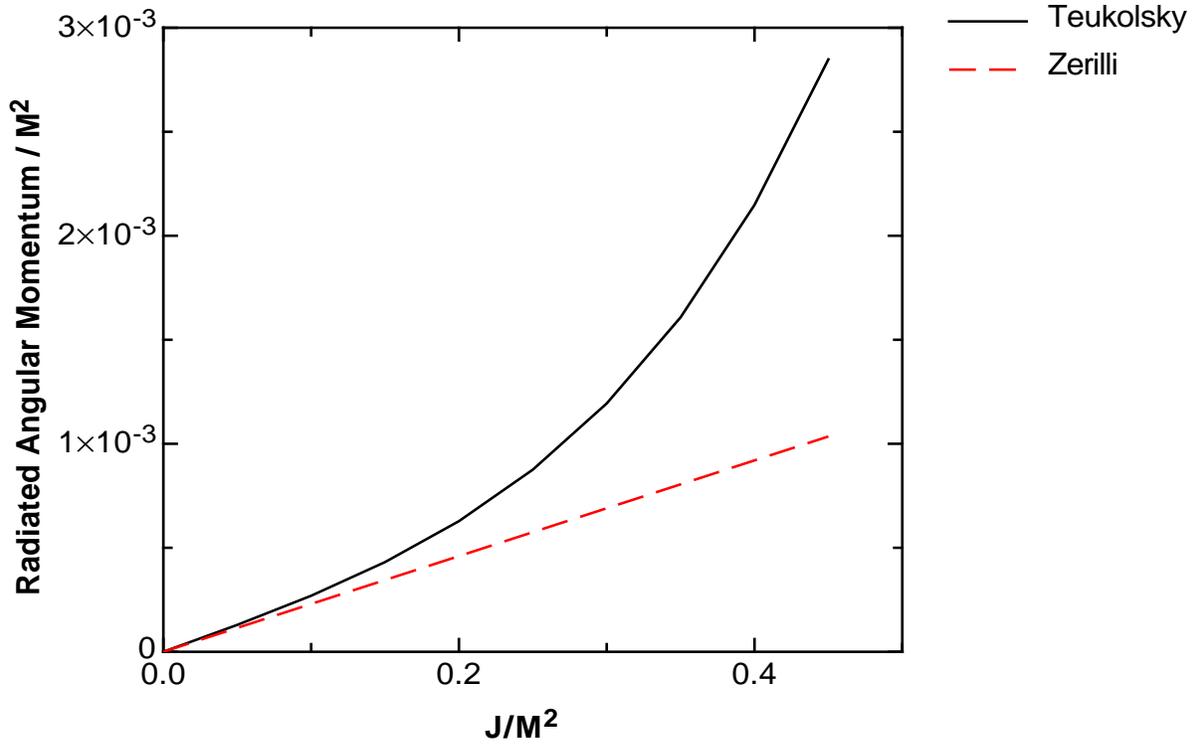}% Here is how to import EPS art
\label{angmom}
\caption{ Radiated angular momenta as a function of total initial angular momentum $r^{*}=25M$. 
Note the good agreement between the two formalisms upto $J/M^{2}=0.2$. All quantities are in units of ADM mass.}
\end{figure*}

\begin{figure*}
\includegraphics[height=10cm,width=16cm]{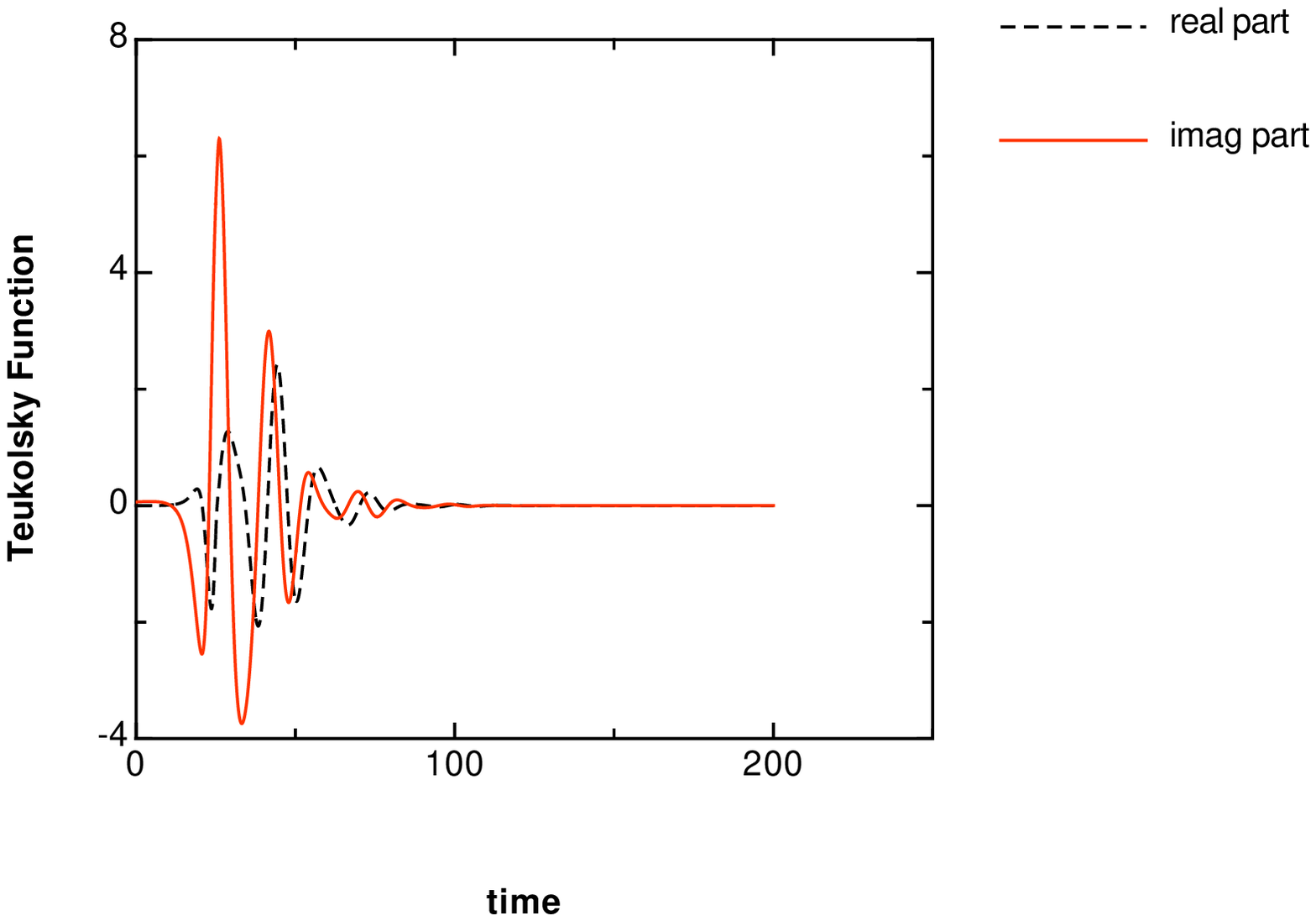}% Here is how to import EPS art
\label{}
\caption{Real and imaginary parts of the $m=2$ mode of the Teukolsky function. All quantities are in units of ADM mass.}
\end{figure*}

\section{Acknowledgments}
The author acknowledges the research support of Southampton College of Long Island 
University. This work is also supported by the National Science Foundation, under grant 
number PHY-0140236. The author also wishes to thank Jorge Pullin for reading a draft 
of the work and providing helpful comments.

\end{document}